\documentclass[preprint, 5p,  times, twocolumn]{elsarticle}
\usepackage{amsmath,amsfonts}
\usepackage{algorithmic}
\usepackage{algorithm}
\usepackage{array}

\usepackage[caption=false, font=normalsize, labelfont=sf, textfont=sf]{subfig}

\usepackage{textcomp}
\usepackage{stfloats}
\usepackage{url}
\usepackage{verbatim}
\usepackage{graphicx}
\usepackage{color}

\usepackage{tikz}
\usetikzlibrary{arrows}
\usetikzlibrary{arrows.meta}
\usetikzlibrary{calc}
\tikzset{block/.style={draw, thick, text width=2cm , minimum height=1.3cm, align=center},   
line/.style={-latex}
}

\journal{Sustainable Energy, Grids and Networks}

\begin{document}

\begin{frontmatter}

\title{Formulation and Experimental Validation of Price-Based Control of Flexible Prosumers in Distribution Grids with the Alternating Direction Method of Multipliers}


\author[inst1]{Plouton Grammatikos}
\author[inst1]{Ali Mohamed Ali}
\author[inst1]{Fabrizio Sossan}

\affiliation[inst1]{organization={University of Applied Sciences of Western Switzerland (HES-SO)},
            addressline={Rue de l'Industrie 21}, 
            city={Sion},
            postcode={1950},
            country={Switzerland}}

            
\begin{abstract}
This paper describes a method for computing price signals for prosumers, incentivizing them to adjust their consumption according to the constraints of the distribution grids to which they are connected, thereby preventing voltage violations and line congestion. The proposed method leverages an interpretation of the Alternating Direction Method of Multipliers (ADMM), which enables the extraction of a price signal to coordinate the operations of prosumers and the distribution grid's constraints while limiting the sharing of sensitive information among them. The method can be used by Distribution System Operators (DSOs) to dynamically adjust a pre-existing retail electricity tariff (e.g., a constant or time-of-use tariff), thereby triggering grid-support actions from prosumers. The method is validated experimentally in a 9-node low-voltage distribution grid laboratory with real components (lines and controllable power converters). The experiments validate the algorithm's performance in terms of convergence and operational efficiency, demonstrating its viability in a real-life setting.
\end{abstract}


\end{frontmatter}

\section{Introduction}

The integration of Distributed Energy Resources (DERs), such as Photo-Voltaic (PV) plants, Electric Vehicles (EVs), and Battery Energy Storage Systems (BESSs) into residential and commercial behind-the-meter settings has allowed passive consumers to turn into prosumers.
Prosumers can leverage the flexibility of their DERs to enhance behind-the-meter performance (e.g., minimizing electricity costs, improving PV self-consumption, and peak shaving), and potentially support grid operations, particularly if they are remunerated. Grid support can involve providing ancillary services to the Transmission System Operator (TSO), such as grid balancing power, or assisting the Distribution System Operator (DSO) with congestion management and voltage control.



Control strategies for flexible prosumers are commonly categorized as direct or indirect control \cite{kosek2013overview}. In direct control, power setpoints are explicitly defined to control resources; in indirect control, an incentive signal is communicated to influence the generation and consumption levels of these resources. More recent definitions of direct and indirect control refer to explicit and implicit flexibility \cite{nouicer2020economics}.
A common form of indirect control is price-based control, where dynamic electricity tariffs incentivize consumption (or generation) shifts under the assumption that consumers aim to minimize costs. Time-of-Use (ToU) electricity tariffs are a typical example of price-based control, implemented to shift demand to off-peak periods (e.g., \cite{filippini2011short, yang2013electricity}). Real-time electricity markets further extend the concept of ToU tariffs by incorporating dynamically adjusted prices that vary based on the available renewable generation (e.g., \cite{wang2015review, nicolson2018consumer}). Compared to other strategies, price-based control is easily understood by prosumers, offering them a clear economic incentive to adapt their consumption. However, existing schemes like ToU and real-time pricing often neglect distribution grid constraints.
As a result, low or high prices may synchronize prosumer behavior in ways that could overload the distribution grid.

In this paper, we address price-based control of prosumers in distribution grids. Specifically, we develop an algorithm that uses the Alternating Direction Method of Multipliers (ADMM, \cite{8186925}) to solve a linearized Optimal Power Flow (OPF) in a distributed way; the algorithm calculates a non-linear price signal capable of steering price-sensitive prosumers' consumption to restore correct operations of the distribution grids in terms of line current and voltage levels. The algorithm is general and can be used to control different types of energy resources without any modifications. In addition to the generalized formulation, we present a use case for voltage control using BESSs and curtailable PV plants. We also propose to handle the uncertainty in the prosumers' load demand and PV generation using a receding horizon control. We finally validate the algorithm in an experimental setting using a test bench distribution grid.

The remainder of this section reviews the state of the art, highlighting parallels with existing methods and outlining this paper's contributions.
In transmission grids, the Locational Marginal Prices (LMP), derived from the Lagrange Multipliers (LMs) of the Optimal Power Flow (OPF) constraints, are used to compute nodal electricity tariffs that reflect transmission constraints and local generation (e.g., \cite{liu2009derivation}).
Studies such as \cite{7524028, 7038107, 10153563, SINGH2010637} extend LMPs to distribution grids by using LMs to adjust electricity prices in the OPF objective, encouraging prosumer demand shifting. Other works \cite{6913579, 9061809} further refine LMs by adopting quadratic cost functions to avoid solution multiplicity. Similarly, we leverage the OPF formulation to compute dynamic tariffs for prosumers, incorporating LMs into a quadratic cost function, as in \cite{6913579, 9061809}, to improve convergence. Compared to these works, we adopt a distributed formulation that separates grid operators from prosumers, preventing the direct exchange of sensitive data.

Existing methods can also be classified depending on the grid model used to compute the LMs, either linearized models (e.g., \cite{7524028}) or approximated non-linear ones (e.g., \cite{8089425, WANG2025101648}), which are more accurate but harder to solve. In our approach, we utilize linearized grid models because, as will be demonstrated in this paper, they lend themselves to a decomposition according to the ADMM paradigm. However, the grid model is improved in each iteration of the algorithm through sequential linearization, which limits the effect of model inaccuracies on the optimal solution and constraint satisfaction.

The use of ADMM for solving OPFs in a distributed manner has been explored in \cite{6748974, gupta2020grid}, while \cite{admm-dispatch} applies ADMM to coordinate the dispatch of curtailable PV plants and BESSs over a time horizon at a single node. The work in \cite{7747961} reviews several forms of ADMM for economic dispatching depending on the grid model (DC or AC OPF). In the context of price-based control, \cite{DONG2024122893} employs a consensus ADMM for peer-to-peer energy trading among prosumers. The grid constraints are incorporated into the LMs, which are computed analytically by the DSO and communicated to the prosumers. Similarly, \cite{8673657, 10402924} utilize the interpretation of the Lagrange multipliers as marginal prices that incentivize the prosumers to change their demand. Additionally, \cite{10659236} highlights ADMM’s privacy-preserving features and shows how it can obfuscate sensitive agent information. 
Compared to existing works, we reinterpret the ADMM formulation as price-based control to drive prosumers' flexibility without explicitly exchanging LMs between participants. This allows the prosumers to interpret the price signals outside the context of ADMM, as they do not need to be aware of the underlying mechanism that computes the LMs.

Other approaches utilize game theory instead of LMPs to control flexible prosumers through pricing, as in \cite{10038110, 8587495}.
Prosumers decide on their strategies through a game, either in coordination with an aggregator aiming to maximize its gain, as in \cite{10038110}, or in coordination with the DSO that aims to satisfy the grid's constraints, as in \cite{8587495}. Compared to our work, the price of the constraints in \cite{8587495} is computed using a dynamic control law, instead of ADMM.

In summary, the key contributions of this paper to the existing state-of-the-art are as follows:
\begin{itemize}
    \item the formulation of a distributed algorithm that solves OPF using ADMM to control various types of flexible resources owned by prosumers without revealing sensitive information to other prosumers or the DSO;
    \item an interpretation of the objective of the distributed OPF as a price signal for generic DERs that accounts for the network constraints of the distribution grid and can be easily interpreted by prosumers without knowledge of the underlying method;
    \item experimental validation of the algorithm for voltage control on a laboratory distribution grid, featuring five prosumers, equipped with BESSs and curtailable PV plants. We also propose receding-horizon and real-time control schemes that complement the distributed OPF to account for grid uncertainties and measurement errors in real hardware. The experiments validate the algorithm's viability for adjusting prosumers' consumption under dynamic tariffs in real conditions.
\end{itemize}

The rest of the paper is organized as follows: Section~\ref{sec:centralized_problem} formulates the centralized problem to be solved by the DSO; Section~\ref{sec:distributed_algorithm} presents the distributed formulation of the problem and provides a novel interpretation of ADMM for price-based control of price-sensitive prosumers; Section~\ref{sec:voltage_control} presents an application example of the proposed algorithm tailored to voltage control, whose experimental validation is then presented in Section~\ref{sec:experimental_validation}. Section~\ref{sec:conclusion} concludes the paper.

\section{Problem Formulation}
\label{sec:centralized_problem}
\subsection{Settings and notation}
We consider $N$ prosumers connected to a low-voltage distribution grid. Each prosumer $i \in \{1,\dots,N\}$ has a set of $M_i$ behind-the-meter energy resources producing or consuming power. Additionally, we consider $K$ discrete time steps, indexed by $t_k \in \{t_1,...,t_K\}$. $\mathbf{x}_{i,t_k}=[p_{i,t_k}, q_{i,t_k}]^\top$ denotes the vector of the total active and reactive power demand of prosumer $i$ at time $t_k$. A negative demand denotes power generation. The full time series for power demand for prosumer $i$ across the $K$ time intervals is denoted by
\begin{align}
    \mathbf{x}_i=[\mathbf{x}_{i,t_1};\mathbf{x}_{i, t_2}; \dots ;\mathbf{x}_{i,t_K}],
\end{align}
where $[;]$ denotes vertical concatenation. $\mathbf{x}_i$ is a $2K\times1$ vector consisting of the total active and reactive power demand for each time step.

Furthermore, the $2KM_i \times 1$ vector
\begin{equation}
       \tilde{\mathbf{x}}_i=[
       \tilde{\mathbf{x}}^1_{i}; \tilde{\mathbf{x}}^2_{i}; \dots;
       \tilde{\mathbf{x}}^{M_i}_{i}
       ] 
\end{equation}
denotes the demand over time of each resource $j \in \{1,2, \dots ,M_i\}$ of prosumer $i$, where $\tilde{\mathbf{x}}^j_{i}$ denotes the power demand of resource $j$ of prosumer $i$ for all times $\{t_1,...,t_K\}$. The sum of the power demand of all resources equals the demand of the prosumer; this reads as:
\begin{align}
    \mathbf{x}_{i} = \sum_{j=1}^{M_i} \tilde{\mathbf{x}}^j_{i}, \forall i \in \{i,...,N\}.
\end{align}

\subsection{Centralized Optimization Problem}
Let the vector $\mathbf{c}'=[c_{t_1}, \dots, c_{t_K}]^\top$ represent the retail electricity price expressed in monetary cost per energy purchased or sold (e.g., in CHF/kWh) for each time interval.
The electricity costs for prosumer $i$ within the period between $t_1$ and $t_K$  can be computed as $(\mathbf{c}')^\top\mathcal{S}\mathbf{x}_i$, where $\mathcal{S}$ is a transformation matrix that extracts the elements corresponding to active power from the vector $\mathbf{x}_i$. Assuming that the DSO's objective is to satisfy the grid's constraints, the problem can be formulated as minimizing the costs of prosumers subject to the resources' and the grid's constraints. By defining $\mathbf{c}=S^\top \mathbf{c}'$, the centralized optimization problem is the following:
\begin{subequations}
    \label{eq:centralized_problem}
    \begin{align}
    \label{eq:aggregated_cost}
    & \min_{\mathbf{x}_i, \tilde{\mathbf{x}}_i, \forall i} && \sum_{i=1}^N \mathbf{c}^\top \mathbf{x}_i \\
    &\text{subject to: } && 
    \label{eq:prosumer_eq_constraints}
    \mathbf{f}_i^{eq}(\tilde{\mathbf{x}}_i) = \mathbf{0}, & \forall i \in \{1,...,N\}\\
    \label{eq:prosumer_ineq_constraints}
    & && \mathbf{f}_i^{ineq}(\tilde{\mathbf{x}}_i) \leq \mathbf{0}, & \forall i \in \{1,...,N\} \\
    \label{eq:prosumer_aggregated}
    & && \mathbf{x}_{i} = \sum_{j=1}^{M_i} \tilde{\mathbf{x}}^j_{i}, & \forall i \in \{1,...,N\}\\
    \label{eq:dso_eq_constraints}
    & && \mathbf{g}^{eq}(\mathbf{x}) = \mathbf{0} \\
    \label{eq:dso_ineq_constraints}
    & && \mathbf{g}^{ineq}(\mathbf{x}) \leq \mathbf{0}
    \end{align}
\end{subequations}
where $\mathbf{f}_i^{ineq}, \mathbf{f}_i^{eq}$ are vector functions representing the equality and inequality constraints, respectively, of prosumer $i$, whereas $\mathbf{g}^{eq}, \mathbf{g}^{ineq}$ are vector functions representing the equality and inequality of the DSO constraints (through load flow equations), respectively. These constraints are a function of the aggregated demand vectors $\mathbf{x}$, which represent the active and reactive power injections into the distribution grid. Note that for Problem~\eqref{eq:centralized_problem} to be convex, the functions $\mathbf{f}_i^{eq}$ must be affine and the $\mathbf{f}_i^{ineq}$ must be convex, which can be enforced by design. However, the convexity of the DSO constraints cannot be guaranteed due to the non-convex characteristics of the power flow equations. In the following section, we consider a linearized grid model, which ensures convexity.

\section{Distributed Algorithm}
\label{sec:distributed_algorithm}
To solve Problem~\eqref{eq:centralized_problem}, it is necessary to centralize measurements and models from both the grid operator and the prosumers into a unified optimization framework. However, this approach is not practical, as it requires both parties to disclose potentially sensitive information, violating the principle of separation of concerns. To overcome this challenge, we reformulate the optimization problem in a distributed manner, inspired by existing literature, and utilize the sharing problem use case formulated with ADMM \cite{8186925}.



\subsection{Separation of the optimization problem}
\label{sec:problem_separation}
To reformulate Problem~\eqref{eq:centralized_problem} into a distributable version, we assume that the constraints \eqref{eq:dso_eq_constraints} and \eqref{eq:dso_ineq_constraints} are either linear or can be linearized. This assumption leads to separable optimization problems for each prosumer, following the method of Dual Decomposition \cite{8186925}. Under this assumption, \eqref{eq:dso_eq_constraints} and \eqref{eq:dso_ineq_constraints} can be expressed as follows:
\begin{subequations}\label{eq:complicatingconstraints}
    \begin{align}
        A^{eq} \mathbf{x} &= \mathbf{b}^{eq} \\
        A^{ineq} \mathbf{x} &\leq \mathbf{b}^{ineq},
    \end{align}
\end{subequations}
where $A^{eq}$ and $A^{ineq}$ are the equality and inequality matrices of the DSO constraints, and $\mathbf{b}^{eq}$ and $ \mathbf{b}^{ineq}$ are the constant terms. 
We then partition the linear equality and inequality matrices column-wise as:
\begin{subequations}
    \begin{align}
        A^{eq}&=\begin{bmatrix} A_1^{eq} & \dots & A_N^{eq} \end{bmatrix} \\
        A^{ineq}&=\begin{bmatrix}A_1^{ineq} & \dots & A_N^{ineq}\end{bmatrix},
    \end{align}
\end{subequations}
where submatrices $A_i^{eq}$ and $A_i^{ineq}$ contain the columns of $A^{eq}$ and $A^{ineq}$ that correspond to prosumer $i$. 
We also introduce the copied vectors $\mathbf{z}_i^{eq}$ and $\mathbf{z}_i^{ineq}$:
\begin{subequations}\label{eq:copiedvectors}
\begin{align}
    \mathbf{z}_i^{eq} &= A_i^{eq} \mathbf{x}_i \\
    \mathbf{z}_i^{ineq} &= A_i^{ineq} \mathbf{x}_i,
\end{align}
\end{subequations}
for all prosumers $i=1,\dots,N$. 
The vectors $\mathbf{z}_i^{eq}, \mathbf{z}_i^{ineq}$ can be interpreted as the contribution of prosumer $i$ to the DSO constraints.

For ease of notation, we introduce the matrices $A_i=[A_i^{eq}; A_i^{ineq}]$ and the vectors $\mathbf{z}_i=[\mathbf{z}_i^{eq}; \mathbf{z}_i^{ineq}]$.
Then, Problem~\eqref{eq:centralized_problem} can be equivalently rewritten as:
\begin{subequations}
    \begin{align}
    \label{eq:total_objective}
    & \min_{\mathbf{x}_i, \tilde{\mathbf{x}}_i, \forall i} && \sum_{i=1}^N  \mathbf{c}^\top \mathbf{x}_i \\
    &\text{subject to: } && \eqref{eq:prosumer_eq_constraints}-\eqref{eq:prosumer_aggregated}, & \forall i \in \{1,...,N\} \\
    \label{eq:prosumer_z_constraints}
    & && A_i \mathbf{x}_i = \mathbf{z}_i, & \forall i \in \{1,...,N\} \\
    \label{eq:dso_z_eq_constraints}
    & && \sum_{i=1}^N \mathbf{z}_i^{eq} = \mathbf{b}^{eq} \\
    \label{eq:dso_z_ineq_constraints}
    & && \sum_{i=1}^N \mathbf{z}_i^{ineq} \leq \mathbf{b}^{ineq}.
\end{align}
\end{subequations}
In the spirit of duality, the constraints~\eqref{eq:prosumer_z_constraints} can be incorporated into a Lagrangian cost function, enabling the cost function to be separated across prosumers. In other words, the total cost function can be expressed as the sum of individual cost functions, one for each prosumer. To enhance the convergence of the algorithm, ADMM employs an augmented Lagrangian cost function, which includes a norm-two penalty for the violations of constraints~\eqref{eq:prosumer_z_constraints}. The augmented Lagrangian for each prosumer $i$ reads as:
\begin{align}
\mathcal{L}_i(\mathbf{x}_i) = \mathbf{c}^\top \mathbf{x}_i + \mathbf{y}_i^\top(A_i\mathbf{x}_i-\mathbf{z}_i)+\frac{\rho}{2}||A_i\mathbf{x}_i-\mathbf{z}_i||_2^2, 
\end{align}
where $\mathbf{y}_i = [\mathbf{y}_i^{eq};\mathbf{y}_i^{ineq}]$ is a vector of LMs, and $\rho$ is a weight of the norm-two penalty term.

\begin{figure}[]
\centering {
\small
\tikzstyle{block} = [rectangle, draw, fill=gray!10,
    text width=15em, text centered, rounded corners, minimum height=3em]

\newcommand\VDIST{7em}

\begin{tikzpicture}[auto]
    \begin{scope}[shift={(-3,0)}]    
        \node[block, text width=5em, anchor=east] (dso) {DSO};
        
        \draw[-{Stealth[scale=1.5]}, gray] ($(dso.north)+(-2.0em,+0.0em)$) arc
        [
            start angle=31,
            end angle=360-31,
            y radius=3em,
            x radius=1.5em
        ];
        
        \node[block, text width=6em, right of=dso, node distance=18em, minimum height=3em] (prosumer_i) {Prosumer $i$};
        
        \draw[-{Stealth[scale=1.5]}, gray] ($(prosumer_i.north)+(2.0em,+0.0em)$) arc
        [
            start angle=0+31,
            end angle=360-31,
            y radius=3em,
            x radius=-1.5em
        ];
        
        \node[block, text width=6em, above of=prosumer_i, node distance=\VDIST] (prosumer_1) {Prosumer 1};
        \node[above of=prosumer_i, node distance=\VDIST/2] () {$\vdots$};
        
        \draw[-{Stealth[scale=1.5]}, gray] ($(prosumer_1.north)+(2.0em,+0.0em)$) arc
        [
            start angle=0+31,
            end angle=360-31,
            y radius=3em,
            x radius=-1.5em
        ];
        
        \node[block, text width=6em, below of=prosumer_i, node distance=\VDIST] (prosumer_N) {Prosumer N};
        \node[below of=prosumer_i, node distance=\VDIST/2] () {$\vdots$};
        
        \draw[-{Stealth[scale=1.5]}, gray] ($(prosumer_N.north)+(2.0em,+0.0em)$) arc
        [
            start angle=0+31,
            end angle=360-31,
            y radius=3em,
            x radius=-1.5em
        ];
        
        \draw[{Stealth[scale=1.5]}-, red] ($(dso.east)+(0.75em,+1.5em)$) -- ($(prosumer_1.west)+(-0.5em,0.5em)$) node [midway,above,xshift=-1em] {$\mathbf{x}^{k+1}_1$};
        \draw[-{Stealth[scale=1.5]}] ($(dso.east)+(0.75em,+0.5em)$) -- ($(prosumer_1.west)+(-0.5em,-0.5em)$) node [midway,below,xshift=3.5em,yshift=1em] {$\mathcal{C}_1^k(\mathbf{x}_1)$};
        
        \draw[-{Stealth[scale=1.5]}] ($(dso.east)+(0.75em,0.30em)$) -- ($(prosumer_i.west)+(-0.5em,0.3em)$) node [midway,above,xshift=0.5em] {$\mathcal{C}_i^k(\mathbf{x}_i)$};
        \draw[{Stealth[scale=1.5]}-, red] ($(dso.east)+(0.75em,-0.30em)$) -- ($(prosumer_i.west)+(-0.5em,-0.3em)$)node [midway,below,xshift=0em] {$\mathbf{x}^{k+1}_i$};

        \draw[-{Stealth[scale=1.5]}] ($(dso.east)+(0.75em,-0.5em)$) -- ($(prosumer_N.west)+(-0.5em,+0.5em)$) node [midway,above,xshift=3.5em, yshift=-0.5em] {$\mathcal{C}_N^k(\mathbf{x}_N)$};
        \draw[{Stealth[scale=1.5]}-, red] ($(dso.east)+(0.75em,-1.5em)$) -- ($(prosumer_N.west)+(-0.5em,-0.5em)$) node [midway,below,xshift=-1em]
        {$\boldsymbol{x}^{k+1}_N$};;
    \end{scope}
    
    \begin{scope}[shift={(-4,-3.5)}]
        \draw[-{Stealth[scale=1.25]}, gray] (0,0) arc
        [
            start angle=0+15,
            end angle=360-15,
            y radius=1.5em,
            x radius=-0.6em
        ];
        \node [text width=5em] at (1.5,-0.1) {Optimization re-iteration} ;
        
        \draw[{Stealth[scale=1.5]}-, red] (2.8,-0.1) -- (3.5,-0.1) node [right] {Demand};
        
        \draw[-{Stealth[scale=1.5]}] (4.8,-0.1) -- (5.5,-0.1) node [right] {Price signal};
    \end{scope}
\end{tikzpicture}
}
\caption{Exchange of information between the DSO and the prosumers during the phases of ADMM. At iteration $k$, The DSO advertises to prosumer $i$ the price signal $\mathcal{C}_i^k(\mathbf{x}_i)$, whose coefficients depend on (i) the sensitivity matrix $A^k_i$ of the DSO constraints with respect to the demand $\mathbf{x}_i$ of the prosumer, (ii) the marginal price $\mathbf{y}^k_i$ of the constraint with respect to $\mathbf{x}_i$, (iii) the contribution $\mathbf{z}^k_i$ of the prosumer in the grid constraints, and (iv) the penalty term $\rho^k$ on the satisfaction of the constraints. The prosumer responds with the optimal decision $\mathbf{x}^{k+1}_i$ that minimizes the price signal subject to the prosumer's constraints.}
\label{fig:information_exchange}
\end{figure}
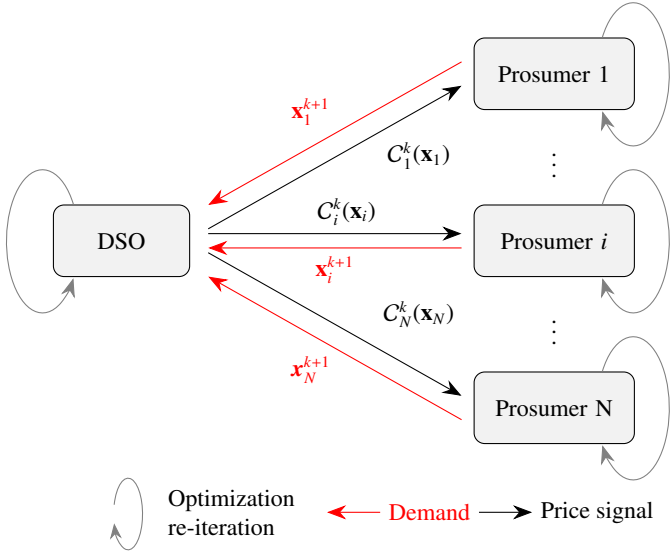

The steps of ADMM to solve Problem~\eqref{eq:total_objective} with the augmented Lagrangian are as follows and are graphically summarized in Fig.~\ref{fig:information_exchange} (see Section~\ref{sec:admm_interpretation_new} for the interpretation of the price signal). With a slight abuse of notation in the following formulation, $k$ in $\mathbf{y}_i^{k+1}$ represents the iteration number rather than an exponent; the same applies to other expressions involving the index 
$k$.

\textbf{Step 0}\; Initialization

\textbf{Step 1}\; Each prosumer $i$ optimizes its own demand $\mathbf{x}_i$ by solving the following optimization problem:
\begin{subequations}
\label{eq:prosumer_objective_admm}
\begin{align}
    \label{eq:prosumer_objective_admm:cost}
    \mathbf{x}_i^{k+1} = &\arg\min_{\mathbf{x}_i} \mathcal{C}^k_i(\mathbf{x}_i) \\
    & \text{subject to:} \eqref{eq:prosumer_eq_constraints}-\eqref{eq:prosumer_aggregated},
\end{align}
where
\begin{equation}
    \label{eq:ci_definition}
    \mathcal{C}_i^k(\mathbf{x}_i) = \mathbf{c}^\top \mathbf{x}_i + (\mathbf{y}_i^k)^\top(A_i^k\mathbf{x}_i-\mathbf{z}_i^k)+\frac{\rho^k}{2}||A_i^k\mathbf{x}_i-\mathbf{z}_i^k||_2^2
\end{equation}
\end{subequations}
is a cost function that can be interpreted as a price signal, as explained in Section~\ref{sec:admm_interpretation_new}. This step can be carried out in parallel by all prosumers.

\textbf{Step 2}\; The DSO collects the results of the optimization from all prosumers ($\mathbf{x}_i^{k+1}$, for all $i$) and updates the copied vectors $\mathbf{z}^{k+1}=[\mathbf{z}^{k+1}_1; \dots; \mathbf{z}^{k+1}_N]$:
\begin{subequations}
\begin{align}
    \mathbf{z}^{k+1} = &\arg\min_{\mathbf{z}_i,\forall i} && \bigg\{ \sum_{i=1}^N (\mathbf{y}_i^k)^\top(A_i^k\mathbf{x}_i^{k+1}-\mathbf{z}_i) + \\
    & && +\frac{\rho^k}{2}||A_i^k \mathbf{x}_i^{k+1}-\mathbf{z}_i||_2^2 \bigg\} \nonumber \\
    & \text{subject to:} && \eqref{eq:dso_z_eq_constraints}-\eqref{eq:dso_z_ineq_constraints}.
\end{align}
\end{subequations}

\textbf{Step 3}\; The DSO updates the LMs by performing the following update:
\begin{align}
    \label{eq:dual_variable_update}
    \mathbf{y}_i^{k+1} = \mathbf{y}_i^k+\rho^k (A_i^k\mathbf{x}_i^{k+1}-\mathbf{z}_i^{k+1}).
\end{align}

\textbf{Step 4}\; The DSO re-linearizes the constraints \eqref{eq:dso_eq_constraints} and \eqref{eq:dso_ineq_constraints} and recomputes the matrices $A_i^{k+1}$ and the vectors $\mathbf{b}^k$ to improve the accuracy of its grid model.

\textbf{Step 5}\; The procedure is repeated.

\subsection{Interpretation of the iterative procedure}
In Step 1, each prosumer minimizes their electricity cost subject to their resource constraints while also accounting for a penalty term imposed by the DSO. This penalty reflects the level of congestion in the distribution grid. During the first iteration, since prosumers optimize their demand independently, violations of grid constraints are likely, especially when very low or high electricity tariffs cause demand to synchronize. In Steps 2 and 3, the grid operator assesses grid constraints and adjusts the LMs, which are then incorporated into the prosumers' optimization problem (Step 1), guiding their consumption toward a feasible grid state through subsequent iterations.
It is important to note that, after convergence, the LM vectors $\mathbf{y}_i$ should be identical for all $i$ as they correspond to the marginal prices of the coupling constraints \eqref{eq:dso_eq_constraints} and \eqref{eq:dso_ineq_constraints} as per dual decomposition theory \cite{8186925}.
Finally, the DSO can re-linearize the constraint matrices $A_i^k$ to reduce the effect of the linearization of $\mathbf{g}^{eq}$ and $\mathbf{g}^{ineq}$. 

\subsection{Interpretation of the formulation as price-based control}
\label{sec:admm_interpretation_new}

Rearranging the terms of the prosumers' costs provides an interpretation of the problem formulation as a price-based control. Rewriting $\mathcal{C}^k_i(\mathbf{x}_i)$ as
\begin{align}
    \label{eq:prosumer_objective_rewriten}
    \mathcal{C}_i^k(\mathbf{x}_i) = 
    &\left[ \mathbf{c}+(A_i^k)^\top\mathbf{y}^k_i+\rho^k(A_i^k)^\top\left(\frac{A_i^k\mathbf{x}_i}{2}-\mathbf{z}^k_i\right)\right]^\top\mathbf{x}_i+\\
    & + \frac{\rho^k}{2}||\mathbf{z}^k_i||_2^2-(\mathbf{y}^k_i)^\top\mathbf{z}^k_i \nonumber
\end{align}
highlights the quadratic nature of the electricity tariff with respect to the power consumption $\mathbf{x}_i$ of prosumer $i$. In particular, its first term
\begin{align}
    \label{eq:consumption_dependent_price}
    \mathbf{c}+(A_i^k)^\top\mathbf{y}^k_i+\rho^k(A_i^k)^\top\left(\frac{A_i^k\mathbf{x}_i}{2}-\mathbf{z}^k_i\right)
\end{align}
is a consumption-dependent electricity tariff that, when multiplied with $\mathbf{x}_i$ in \eqref{eq:prosumer_objective_rewriten}, yields an electricity cost. The second term
\begin{align}
    \label{eq:price_signal_fixed}
    \frac{\rho^k}{2}||\mathbf{z}^k_i||_2^2-(\mathbf{y}^k_i)^\top\mathbf{z}^k_i
\end{align}
is a fixed cost paid by the prosumers regardless of their demand (it does not depend on $\mathbf{x}_i$), and can thus be interpreted as a service fee for grid connection. Note that after ADMM has converged, the values of $\rho^k$, $A_i^k$, $\mathbf{y}_i^k$, and $\mathbf{z}_i^k$ are fixed, so the coefficients of the final costs $\mathcal{C}_i^k(\mathbf{x}_i)$ advertised to the prosumers are also fixed.

After ADMM convergence, each prosumer minimizes the cost in \eqref{eq:ci_definition} (equivalently \eqref{eq:prosumer_objective_rewriten}), subject to their constraints, to compute power setpoints that minimize their cost while satisfying the DSO’s constraints. This means the DSO can control the prosumers by adjusting the perceived prices instead of setting explicit power setpoints.
It is worth noting that, given the form of \eqref{eq:prosumer_objective_rewriten}, the DSO only needs to advertise the coefficients of $\mathbf{x}_i$ to the prosumer, thus hiding sensitive information about the grid, such as the matrices $A_i$ or the LMs $\mathbf{y}_i$. 

Knowing the price $\mathbf{c}$, the prosumers can compute their compensation for providing flexibility to the DSO. Let $\hat{\mathbf{x}}_i$ denote the demand of prosumer $i$ that minimizes $\mathbf{c}^\top \mathbf{x}_i$ subject to the prosumer's constraints, and $\mathbf{x}_i^*$ represent the solution of the distributed problem at convergence. Then, their compensation could be calculated as follows:
\begin{align}
    \text{compensation}_i = \mathbf{c}^\top (\mathbf{x}_i^*-\hat{\mathbf{x}}_i).
\end{align}
Compensation is non-negative by construction because the inclusion of grid constraints increases the prosumers' costs.

Finally, it should be noted that the value of the penalty term $\rho$ does not affect the optimality condition. This can be derived from the fact that after the algorithm has converged ($k \rightarrow \infty$), the condition $A_i\mathbf{x}_i^*=\mathbf{z}_i$ should hold for all $i$. Therefore, the cost \eqref{eq:ci_definition} evaluated at $\mathbf{x}^*_i$ is $\mathbf{c}^\top\mathbf{x}^*_i$, which is independent of $\rho$. However, the condition $A_i\mathbf{x}_i^*=\mathbf{z}_i$ is only satisfied provided that the LMs have fully converged. As we will verify in Section~\ref{sec:admm_convergence}, a sufficiently large value of $\rho$ is needed to guarantee optimality in a finite number of iterations.

\subsection{Convergence}
\label{sec:convergence}
The convergence of the algorithm is measured by the norm of the primal and dual residuals, defined as:
\begin{subequations}
    \begin{align}
        r_i^{k+1} &= ||A_i\mathbf{x}_i^{k+1}-\mathbf{z}_i^{k+1}||_2 \\
        s_i^{k+1} &= \rho^k||\mathbf{z}_i^{k+1}-\mathbf{z}_i^k||_2.
    \end{align}
\end{subequations}
The iterations halt when both norms are smaller than a given tolerance for all prosumers. The tolerances are computed dynamically according to the guidelines proposed in \cite{8186925}. The penalty term $\rho$ is also computed dynamically with each iteration according to the method proposed in \cite{rho-adaptive}, as follows:
\begin{equation}
    \label{eq:adaptive_rho}
    \rho^{k+1}=
    \begin{cases}
        \tau^{incr} \rho^k & r_i^{k+1}>\mu s_i^{k+1}\\
        \dfrac{\rho^k}{\tau^{decr}} & r_i^{k+1}<\mu s_i^{k+1}\\
        \rho^k & \text{otherwise.}
    \end{cases}
\end{equation}
If the primal residual norm is larger than the dual residual norm by a given factor $\mu$, then we increase the penalty $\rho^k$ by a factor of $\tau^{incr}$ to bring the coupling constraints closer to the feasible region. If, on the other hand, the dual residual norm is too large, we decrease $\rho^k$ by a factor $\tau^{decr}$ to help the convergence of the copied variables and the LMs\footnote{Note that in Eq.~\eqref{eq:dual_variable_update} $\rho^k$ is the step size of the dual variable update.}.

Another metric of convergence is the maximum linearization error of the DSO's constraints, given by
\begin{equation}
    \label{eq:linearization_error}
    r_{lin}^{k+1} = ||\mathbf{g}(\mathbf{x}^{k+1})-(A^{k}\mathbf{x}^{k+1}-\mathbf{b}^k)||_{\infty},
\end{equation}
where $\mathbf{g}=[\mathbf{g}^{eq};\mathbf{g}^{ineq}]$ and $\mathbf{b}^k=[\mathbf{b}^{eq,k};  \mathbf{b}^{ineq,k}]$. To avoid unnecessary recomputations of the matrix $A^k$ and the vector $\mathbf{b}^k$, we compute them at the end of each iteration (Step 4) only if \eqref{eq:linearization_error} is above a given tolerance.

\subsection{Handling uncertainty in the prosumers' demand} \label{sec:uncertainty}

The method presented so far assumes that the behavior of prosumers is known in advance through point predictions. However, uncertainties arising from stochastic load demand and solar irradiance may lead to grid constraint violations or exacerbate existing ones, causing the DSO to overcompensate prosumers for shifting their demand more than necessary.

To mitigate the impact of these uncertainties and to enable the experiments described in Section~\ref{sec:experimental_validation}, the proposed optimization is applied in a receding-horizon fashion. In this approach, the price signal is periodically recomputed based on newly available information that gradually reveals uncertainty. Prosumers can then re-optimize their responses in real time using more refined forecasts to ensure that their decisions remain close to the optimal ones.
Although the proposed framework can accommodate other types of forecasts (such as prediction intervals), the explicit inclusion of uncertainty will be addressed in future work.

In Section~\ref{sec:voltage_control}, we show how to handle uncertainty in a specific use case where flexibility is provided by behind-the-meter BESSs and curtailable PV plans owned by prosumers. Specifically, in Section~\ref{sec:mpc}, we explain how to run a receding horizon optimization to recompute the price signals using updated forecasts of the prosumers' demands and the PV production. In Section~\ref{sec:rt_control}, we formulate a real-time (RT) control to compute the power setpoints implemented by the prosumers in response to the updated price signals.

\section{Application Example: Voltage Control}
\label{sec:voltage_control}

The formulation in the previous section is general and can be adapted to various energy resources and DSO objectives, provided that the sharing constraints are either linear or can be linearized. In this section, we demonstrate its application in a scenario where prosumer flexibility is offered to the DSO for voltage control.

\subsection{Prosumers constraints and objective}
Prosumers are equipped with a BESS with a rated power of $s^{b, max}_i$ (kVA) and an energy capacity of $E^b_i$ (kWh). For simplicity in the following formulation, it is assumed that the BESS has unitary round-trip efficiency and that it can provide active and reactive in the limit imposed by the rated power. Under these assumptions, the state-of-charge (SoC) of the BESS of prosumer $i$ at time $t$ can be approximated as:
\begin{subequations}
\label{eq:bess:model}
\begin{align}
    \label{eq:bess:soc_cons}
    SoC_{i,t+1} = SoC_{i,t_1} - \frac{\Delta T}{E^b_i} \sum_{\tau=t_1}^{t} p^b_{i,\tau}
\end{align}
where $SoC_{i,t_1}$ is an initial SoC, $p^b_{i,t}$ is the BESS discharging power at time $t$ (negative if charging), and $\Delta T$ is the timestep of the day-ahead optimization; the battery SoC must remain within configurable $SoC^{min}_i$ and $SoC^{max}_i$ limits:
\begin{align}
SoC^{min}_i \leq SoC_{i,t+1} \leq SoC^{max}_i.    
\end{align}
The active and reactive power limits of the BESS are approximated using inner box constraints derived from the PQ circular capability curve of the converter and are:
\begin{align}
-{s_i^{b,max}} &\leq \sqrt{2} \cdot  p^b_{i,t} \leq {s_i^{b,max}} \\
-{s_i^{b,max}} &\leq \sqrt{2} \cdot q^b_{i,t} \leq {s_i^{b,max}}
\end{align}
\end{subequations}
where $q^b_{i,t}$ is the reactive power of the converter.

Prosumers are also equipped with a PV installation, whose power output might be curtailed if necessary. Let the active power generation potential at time $t$ be denoted by $p^{pv, max}_{i,t}$. The PV generation $p^{pv}_{i,t}$ should satisfy the condition:
\begin{subequations}
\label{eq:cpv:model}
\begin{align}
0 \leq p^{pv}_{i,t} \leq p^{pv, max}_{i, t}
\end{align}
where the $p^{pv, max}_{i, t} - p^{pv}_{i,t}$ represents the curtailment action. We assume that the PVs only provide active power flexibility so we assume that the reactive power is zero for all timesteps
\begin{equation}
    q^{pv}_{i,t} = 0
\end{equation}
\end{subequations}
Let the vectors $\mathbf{p}_i^l, \mathbf{q}_i^l$
denote the active and reactive load demand of prosumer $i$ for all time steps. Similarly, the BESS active and reactive power and the PV production for all time steps are collected in the vectors $\mathbf{p}_i^b, \mathbf{q}_i^b$ and $\mathbf{p}_i^{pv}, \mathbf{q}_i^{pv}$. 
The cost-optimal trajectories of the BESS power and PV curtailment are given by solving the following optimization problem:
\begin{subequations}
\label{eq:prosumer_problem_application}
\begin{align}
\label{eq:prosumer_cost_application}
\min_{\mathbf{p}_i^{pv},\mathbf{p}_i^b, \mathbf{q}_i^{pv},\mathbf{q}_i^b, , \mathbf{x}_i} \mathcal{C}^k_i(\mathbf{x}_i)
\end{align}
subject to:
\begin{align}
& \mathbf{x}_i=(\mathbf{p}_i^l-\mathbf{p}_i^{pv}-\mathbf{p}_i^b, \mathbf{q}_i^l-\mathbf{q}_i^{pv}-\mathbf{q}_i^b) \\
&\text{Battery SOC and power constraints \eqref{eq:bess:model} } && \forall t \\
&\text{Curtailable PV model \eqref{eq:cpv:model} } && \forall t,
\end{align}
\end{subequations}
where $\mathcal{C}^k_i(\mathbf{x}_i)$ is given by \eqref{eq:prosumer_objective_rewriten} and $k$ is the iteration number of ADMM at convergence.

\subsection{DSO Constraints}
The objective of the DSO in this use case is to keep the nodal voltage magnitudes within given bounds $v_{min}, v_{max}$. We compute the voltages using a linearized grid model:
\begin{equation}
    \mathbf{v}=\mathbf{v}^*+K_{v,p}(\mathbf{p}-\mathbf{p}^*)+K_{v,q}(\mathbf{q}-\mathbf{q}^*)
\end{equation}
where $\mathbf{v}$ is the vector of nodal voltage magnitudes for all timesteps and $\mathbf{p}, \mathbf{q}$ are the vectors of the nodal active and reactive power injections, respectively. $K_v$ is the matrix of the voltage sensitivity coefficients \cite{sensitivities}, which is computed around a given set of setpoints $[\mathbf{p}^*; \mathbf{q}^*]$. $\mathbf{v}^*$ are the nodal voltages when the power injections are $[\mathbf{p}^*; \mathbf{q}^*]$. In our implementation, the vector $[\mathbf{p}^*; \mathbf{q}^*]$ is the solution of the individual prosumers problems given by \eqref{eq:prosumer_problem_application}.

The DSO constraints can be written as:
\begin{subequations}
    \begin{equation}
        -[K_{v,p}, K_{v,q}] [\mathbf{p};\mathbf{q}] \leq \mathbf{v}^* - v_{min} - K_{v,p} \mathbf{p}^* - K_{v,q} \mathbf{q}^*
    \end{equation}
    \begin{equation}
        [K_{v,p}, K_{v,q}] [\mathbf{p};\mathbf{q}] \leq v_{max} -\mathbf{v}^* + K_{v,p} \mathbf{p}^* + K_{v,q} \mathbf{q}^*
    \end{equation}
\end{subequations}
from which we can compute the matrix $A^{ineq}$ and the vector $\mathbf{b}^{ineq}$. The sensitivity coefficients can be re-computed in each iteration of ADMM to minimize the linearization error (Step 4 of the algorithm in Section~\ref{sec:problem_separation}).

The DSO also needs to keep the power factor at the slack bus higher than a desired value.
However, power factor constraints are typically nonconvex, so instead we simply limit the reactive power at the slack bus by a tight upper bound:
\begin{subequations}
\begin{align}
    \label{eq:slack_q}
    |q_{s,t}| \leq q_s^{max}, \forall t,
\end{align}    
where $q_s^{max} \ll s_s^{max}$ and $s_s^{max}$ is the maximum apparent power at the slack bus. Then, the active power limit is defined as
\begin{equation}
    |p_{s,t}| \leq \sqrt{(s_s^{max})^2-(q_s^{max})^2}
\end{equation}
\end{subequations}

\subsection{Receding horizon formulation}
\label{sec:mpc}
In practice, the load profiles, PV generation, and voltage at the slack bus are not known in advance and must be forecasted to effectively schedule energy storage operations as in problem formulation \eqref{eq:prosumer_problem_application}. To mitigate the effect of these uncertainties, we update the price signals using a receding horizon strategy, as was explained in Section~\ref{sec:uncertainty}.

Day-ahead forecasts of the load profiles and the maximum PV production are denoted as $\hat{p}_{i,t}^l$ and $\hat{p}_{i,t}^{pv,max}$ for $i \in \{1,...,N\}$ and $t \in [0, 24h]$, respectively.
The day before operation, each prosumer solves \eqref{eq:prosumer_problem_application} with these forecast values, and the DSO computes an appropriate price signal for each prosumer for the whole day with the algorithm of Section~\ref{sec:distributed_algorithm}; then, the algorithm is rerun during the day every $T_1$~s (e.g., 10-15 minutes) to compute updated price signals for the prosumers using the most recent measurements of the BESSs SoC and updated forecasts of the PV generation and the load profiles. 
At a given time $\tau$, the updated price signal refers to the time interval $[\tau, 24h]$ (shrinking horizon). This control layer will be referred to as the intra-day MPC.

\subsection{Real-time (RT) control}
\label{sec:rt_control}
Since loads and PV generation can fluctuate between two MPC cycles due to short-term variations, prosumers may not be able to implement the PV and BESS profiles computed during the coordination, as doing so could violate grid constraints. Instead, each prosumer should adapt their demand to accommodate load variations, ensuring that the price signal sent by the DSO is always minimized. To achieve this, we propose an RT control layer for each prosumer.

We assume that the intra-day MPC is executed at times $k=\{0, T_1, 2T_1,...\}$, whereas the real-time control is executed at times $\kappa=\{0, T_2, 2T_2,...\}$, where $T_2$ is the computational cycle of the real-time control (e.g., 30~s). At time $\kappa$, the real-time control of prosumer $i$ minimizes the difference between the aggregated demand of the prosumer and the target computed by the intra-day MPC. The problem solved by the real-time control is the following
\begin{subequations}
\label{eq:rt_problem}
\begin{align}    
    \label{eq:rt_objective}
    & \min_{p^b_{i,\kappa}, q^b_{i,\kappa}} && \left\{ (p_{i,\kappa}-p^{target}_{i,\kappa})^2+(q_{i,\kappa}-q^{target}_{i,\kappa})^2 \right\} \\
    & \text{s.t.:}
    && p_{i,\kappa} = \tilde{p}^l_{i,\kappa}-\tilde{p}^{pv}_{i,\kappa}-p^b_{i,\kappa}, \\
    & && q_{i,\kappa} = \tilde{q}^l_{i,\kappa}-\tilde{q}^{pv}_{i,\kappa}-q^b_{i,\kappa}, \\
    & && -\frac{s_i^{b,max}}{\sqrt{2}} \leq p^b_{i,\kappa} \leq \frac{s_i^{b,max}}{\sqrt{2}}, \\
    & && -\frac{s_i^{b,max}}{\sqrt{2}} \leq q^b_{i,\kappa} \leq \frac{s_i^{b,max}}{\sqrt{2}}, \\
    \label{eq:rt_soc}
    & && SoC^{min}_i \leq SoC_{i,\kappa}-\frac{\Delta T}{E^b_i} p^b_{i,\tau} \leq SoC^{max}_i,
\end{align}
\end{subequations}
where $p_{i,\kappa}^{target}, q_{i,\kappa}^{target}$ are the target active and reactive power computed by the prosumer with the intra-day MPC at time $k=\lfloor \kappa T_2/T_1 \rfloor T_1$; $\tilde{p}^l_{i,\kappa},\tilde{q}^l_{i,\kappa}$ is the short-term forecast of the load computed at time $\kappa$ for $\kappa+1$; $\tilde{p}^{pv}_{i,\kappa}, \tilde{q}^{pv}_{i,\kappa}$ is the power of the PV computed by using the forecast of the irradiance for time $\kappa+1$ and the curtailment percentage that was computed by the intra-day MPC at $k=\lfloor \kappa T_2/T_1 \rfloor T_1$. Therefore, we consider that the PV curtailment is given by the MPC, and we compute the power of the BESS to follow the target aggregated setpoint computed by the MPC in the previous cycle. If the computation of the MPC has not finished by time $\kappa$, we use as target setpoint the output of the MPC for time $k=\lfloor \kappa T_2/T_1 \rfloor T_1$ that was computed in the previous cycle.

\section{Results}
\label{sec:experimental_validation}

This section validates the performance of the proposed coordination algorithm in two steps. First, we solve the day-ahead optimization problem and analyze the performance and convergence of the algorithm via simulations. Second, we perform an experimental validation to confirm the method’s practical applicability in the presence of the system's uncertainties. Sections~\ref{sec:setup} and \ref{sec:implementation} detail the experimental setup and the algorithm's parameters, respectively. The simulation results of the day-ahead problem are presented in Sections~\ref{sec:day_ahead} and \ref{sec:costs}, whereas Section~\ref{sec:admm_convergence} analyzes the convergence performance of ADMM when solving the day-ahead problem. Finally, Section~\ref{sec:experiments_results} presents the results of the experimental validation obtained by implementing the receding horizon and RT controls in real hardware.

\subsection{Use case and experimental setup} \label{sec:setup}
\begin{figure}[]
    \centering
    \includegraphics[width=\linewidth]{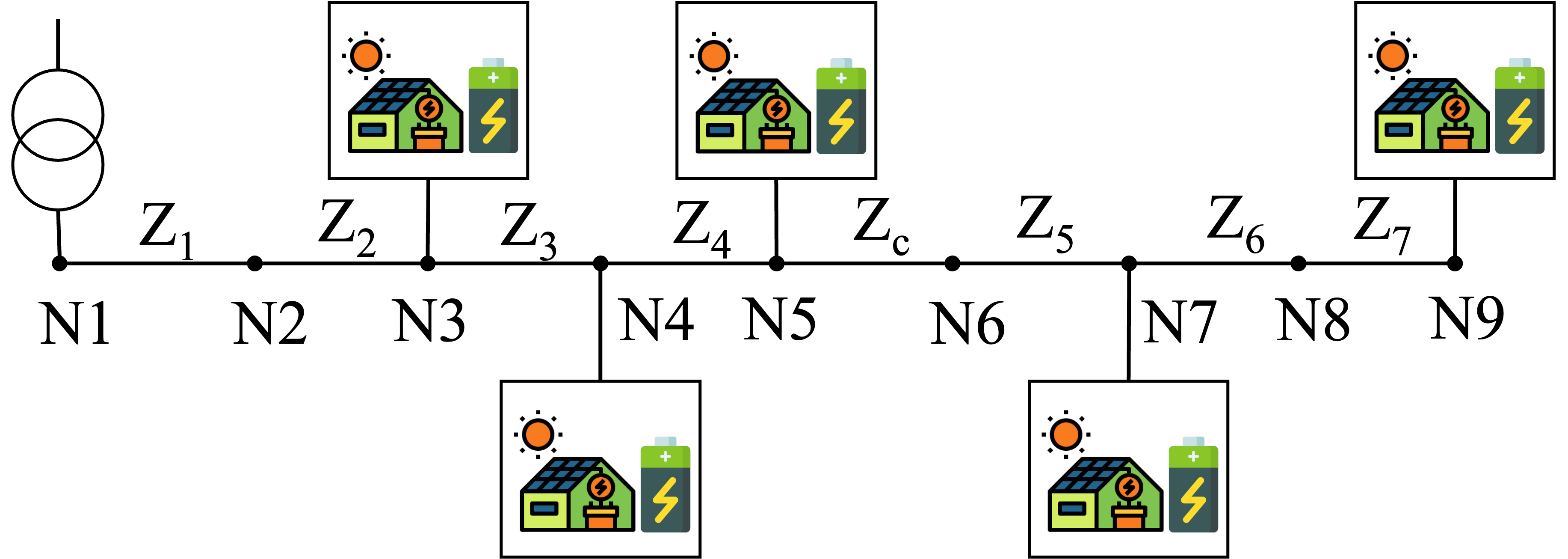}
    \caption{Grid topology for the experimental validation. The emulated prosumers are connected to nodes N3, N4, N5, N7, and N9. N1 is the connection point with the upper grid.}
    \label{fig:topology}
\end{figure}

We consider a DSO aiming to control prosumers via price signals for voltage regulation, as described in the previous section. The proposed method is validated through experiments on a low-voltage, three-phase, radial distribution grid implemented in the Gridlab laboratory at HES-SO Valais in Sion (Switzerland). The grid comprises nine nodes interfacing loads and DERs, along with one additional node that connects to the higher-level grid. Loads and DERs at each node are emulated using 4-quadrant AC/DC back-to-back power converters, each supplied by a separate power line and rated at 15~kVA. The system setup is shown in Fig.~\ref{fig:topology}; node N1 is connected to the external grid and serves as the system's slack bus. Nodes N2, N6, and N8 have no power injections, whereas the remaining nodes interface prosumers, as described later in this subsection.

The nodes are interconnected by either real lines or lines emulated with discrete components based on their pi-model. In particular, Impedances $Z_2$-$Z_4$ and $Z_6$-$Z_7$ are emulated to model 500-m-long aerial lines; $Z_1$ and $Z_5$ are each comprised of an emulated line and a real 60~m cable connected in series.
Table~\ref{tab:line_parameters} shows the line parameters per phase, which have been fine-tuned compared to their nominal values using a numerical procedure based on measurements \cite{line_parameters}. The series impedances for those nodes without demand are combined.
The grid is monitored using SICAM measurement units from Siemens, which provide RMS voltage values at all nodes, line currents, and active and reactive power injections.

\begin{table}[h!]
    \renewcommand{\arraystretch}{1.3}
    \caption{Line parameters per phase}
    \centering
    \begin{tabular}{c|c|c}
         \bf Line & \bf Resistance (m$\Omega$) & \bf Reactance (m$\Omega$)  \\
         \hline
         $Z_1+Z_2$ & 585.1 & 305.7\\
         $Z_3$ & 194.1 & 161.3\\
         $Z_4$ & 216.1 & 177.8\\
         $Z_c$ & 200 & 0 \\
         $Z_5$ & 597.0 & 296.2\\
         $Z_6+Z_7$ & 194.4 & 166.4
    \end{tabular}
    \label{tab:line_parameters}
\end{table}

Nodes N3–N5, N7, and N9 each interface one prosumer, equipped with a BESS, a curtailable PV plant, and inflexible power demand. In the experiments, these components are simulated in software, and their power outputs are reproduced by load emulators to assess their impact on the distribution grid. The power and energy ratings of the prosumers’ assets are listed in Table~\ref{tab:prosumer_ratings}.
Prosumers’ load profiles are based on real measurements from office buildings. We use ten weekdays of data at 30~s resolution: five serve as day-ahead forecasts, and the remaining five as realized loads. For solar irradiance, we use a sunny-day time series as the forecast and assume the realized irradiance is 90\% of the forecast to model uncertainty. The BESS SoC is simulated using a simple power-integral model with unit efficiency.

\subsection{Implementation of the algorithms}
\label{sec:implementation}
The day-ahead dispatching, the intra-day MPC, and the real-time control described in Section~\ref{sec:voltage_control} are implemented in MATLAB and solved with Gurobi. 
The timestep of the day-ahead and MPC decisions is 10~min; the receding horizon optimization is re-executed every 10~min. The real-time control cycle is 30~s. The voltage limits are $v_{min}=0.9$~pu and $v_{max}=1.05$~pu at all nodes. The maximum reactive power at the slack bus is set to $q_0^{max} = 0.1s_0^{max}$, which gives a power factor of over 0.99 when the exported active power to the upper grid is maximum.

Concerning the convergence of ADMM, the initial value of the penalty term is $\rho^0=1$. The $\rho$ adaptation factors defined in Section~\ref{sec:convergence} were chosen empirically by simulations to avoid significant oscillations of the ADMM iterations and are $\tau^{incr}=\tau^{decr}=1.01$ and $\mu=10$. The choice of this value will be justified in Section~\ref{sec:admm_convergence}. Finally, if the primal and dual residuals of the MPC decisions have not converged to the required tolerance within 60~s, then the coordination stops and the price signals are updated based on the last ADMM iteration.

\begin{table}[]
    \caption{Power and Energy Ratings of Prosumers}
    \renewcommand{\arraystretch}{1.3}
    \centering
    \begin{tabular}{l|l}
         \bf Parameter & \bf Value \\
         \hline
         Max Load & 2.5 kW \\
         Max PV power & 5 kW \\
         Rated BESS power & 2.5 kVA \\
         BESS energy capacity & 2.5 kWh\\
    \end{tabular}
    \label{tab:prosumer_ratings}
\end{table}

\subsection{Day-Ahead dispatch}
\label{sec:day_ahead}

\begin{figure}[t!]
  \centering
  \subfloat[]{%
       \includegraphics[width=1\linewidth]{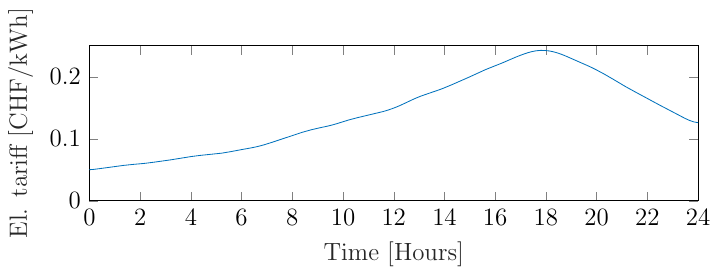}}
  \\
  \subfloat[]{%
        \includegraphics[width=1\linewidth]{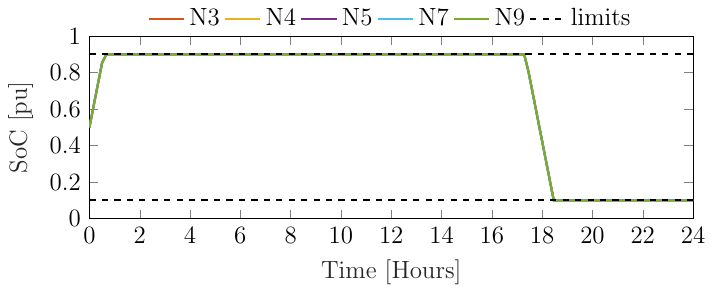}}
  \\
  \subfloat[]{%
        \includegraphics[width=1\linewidth]{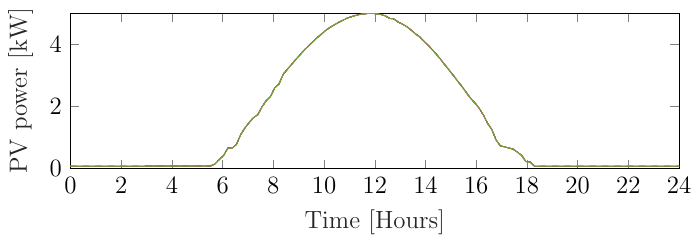}}
  \\
  \subfloat[]{%
        \includegraphics[width=1\linewidth]{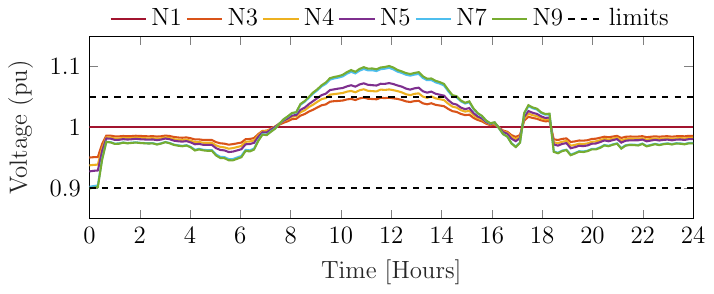}}
  \caption{Simulation results with no DSO-prosumers coordination. (a) Electricity tariff, (b) SoC of the BESSs, (c) PV production, and (d) nodal voltage magnitudes, which violate the upper limit in the central part of the day.}
  \label{fig:results_no_coord}
\end{figure}

We first solve the day-ahead optimization problem using forecasts of the electricity price, the prosumers' loads, and the solar irradiance to schedule the operations of DERs (BESS and PV).
As an initial simulation, the day-ahead optimization problem is solved without coordination among prosumers and DSO. Results are visible in Fig.~\ref{fig:results_no_coord}: the BESSs charge to their maximum capacity early when electricity prices are low and discharge in the evening when prices are higher, in line with their objective of reducing electricity costs. Meanwhile, the PV systems operate at their full potential, avoiding curtailment, as doing so would increase grid import and, consequently, electricity costs. However, violations of voltage constraints occur throughout the day due to the high charge/discharge rates of the BESSs and the elevated PV production levels. This reinforces the motivation of this paper that seeking coordination among DSO and prosumers is essential.


\begin{figure}[t!]
  \centering
  \subfloat[]{%
       \includegraphics[width=1\linewidth]{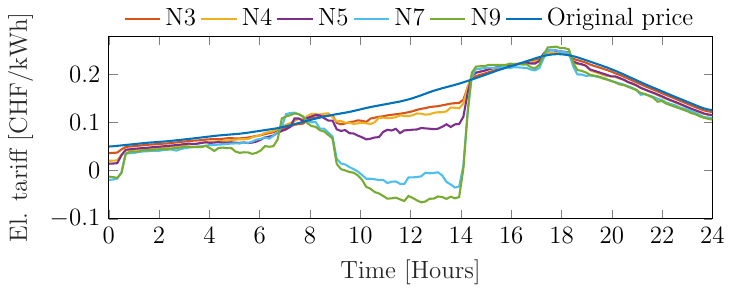}}
  \\
  \subfloat[]{%
        \includegraphics[width=1\linewidth]{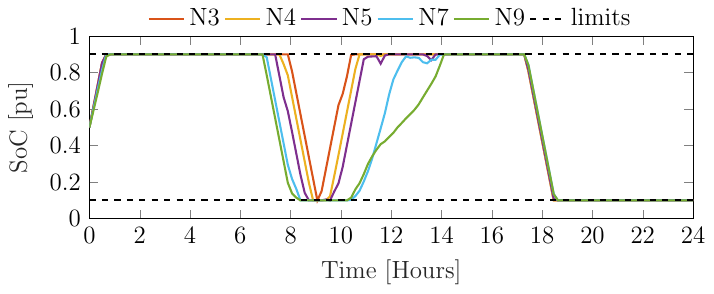}}
        \label{fig:reults_coord}
  \\
  \subfloat[]{\label{fig:coord_pv}
        \includegraphics[width=1\linewidth]{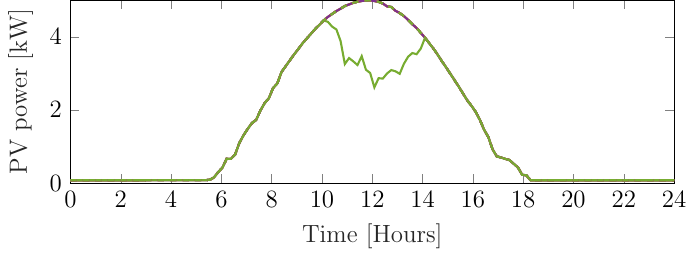}}
  \\
  \subfloat[]{%
        \includegraphics[width=1\linewidth]{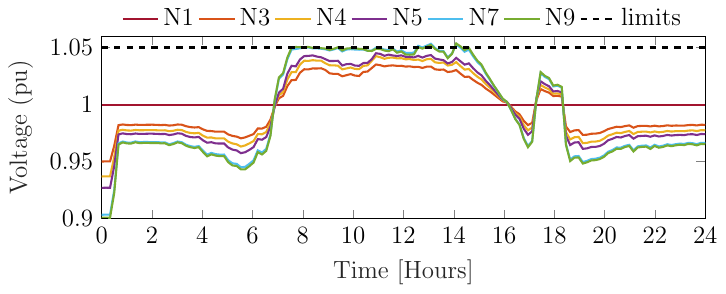}}
  \caption{Simulation results with DSO-prosumers coordination. (a) Electricity tariff, now different for each consumer, (b) SoC of the BESSs, (c) PV production, showing one unit being curtailed, and (d) nodal voltage magnitudes, all of which now fall within limits.}
  \label{fig:results_coord}
\end{figure}

The results with coordination are shown in Fig.~\ref{fig:results_coord}. The BESSs now sell their energy early, allowing them to absorb energy from the PVs during the middle of the day to avoid violating grid constraints. We also observe that the PV generation of the prosumer at node N9 (the furthest from the grid connection point) is curtailed. This results from the BESS's insufficient power and energy regulation capacity to maintain feasible grid constraints. More specifically, Fig.~\ref{fig:results_coord}(b) shows that all SoC levels reach saturation before the end of the curtailment action, indicating that the BESSs lack residual flexibility to prevent the curtailment of PV generation. Consequently, reducing PV generation becomes the last resort to avoid violations of grid constraints.

\subsection{Evolution of the price signals and operation costs} \label{sec:costs}
The plot in the top panel of Fig.~\ref{fig:results_coord} shows the consumption-dependent price of Equation~\eqref{eq:consumption_dependent_price} evaluated at the optimal demand $\mathbf{x}_i=\mathbf{x}_i^*$ for each prosumer $i$ and compares it to the original electricity price $\mathbf{c}$. When there are no binding grid constraints, the price signals provided by the DSO follow the original electricity price. However, when grid constraints are active, the advertised price diverges from the original price. During the central part of the day, when PV generation leads to overvoltages, the DSO advertises a negative price to incentivize prosumers to increase their consumption.

The first two columns of Table~\ref{tab:cost} compare the operational costs of all prosumers with and without DSO coordination. Negative costs indicate a monetary gain. When grid constraints are considered, the monetary gains of all prosumers decrease, as the limitations of the distribution grid prevent them from fully benefiting from the most favorable electricity tariffs. The third column of Table~\ref{tab:cost} shows the price difference; one can hypothesize that the DSO compensates prosumers for this price discrepancy, offering remuneration for their flexibility.

\begin{table}[]
    \caption{Cost of operation in CHF}
    \centering
    \renewcommand{\arraystretch}{1.3}
    \begin{tabular}{c|c|c|c}
         \bf Prosumer & \bf Without coord. & \bf  With coord. & \bf Difference\\
         \hline
         N3 & -0.28 & -0.25 & 0.03\\
         N4 & -0.96 & -0.92 & 0.04\\
         N5 & -0.72 & -0.67 & 0.05\\
         N7 & -0.59 & -0.50 & 0.09\\
         N9 & -0.85 & 0.07 & 0.92\\
         \hline
         Total & -3.39 & -2.26 & 1.13
    \end{tabular}
    \label{tab:cost}
\end{table}

\subsection{Convergence of ADMM and impact of the value of $\rho$}
\label{sec:admm_convergence}

\begin{figure}[t]
    \centering
    \includegraphics[width=\linewidth]{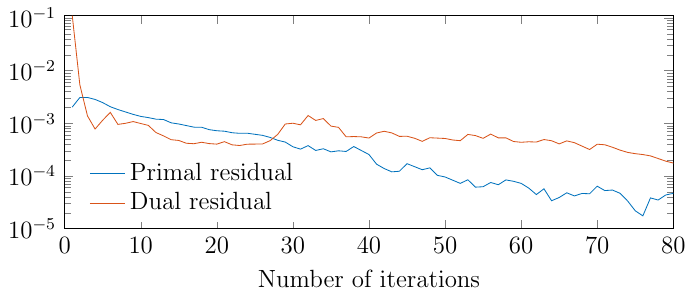}
    \caption{Primal and dual residuals as a function of the number of iterations.}
    \label{fig:convergence}
\end{figure}

Regarding the convergence of ADMM, Fig.~\ref{fig:convergence} shows the norm of the primal and dual residuals (maximum over all prosumers) as a function of the number of iterations. After 80 iterations, both residuals are at around $10^{-4}$, which is sufficient to achieve optimality and satisfy the constraints, as was verified in the previous section. 

\begin{figure}[t]
    \centering
    \includegraphics[width=\linewidth]{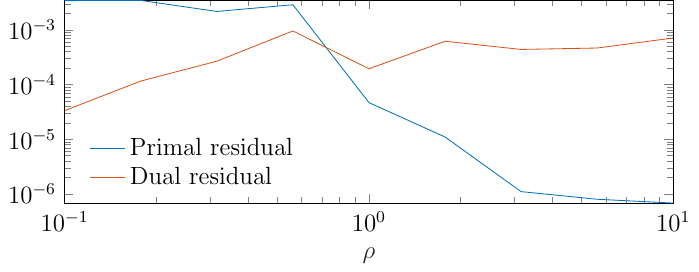}
    \caption{Effect of the parameter $\rho$ on the convergence.}
    \label{fig:residuals_rho}
\end{figure}

As seen in Section~\ref{sec:distributed_algorithm}, $\rho$ has a dual role: as a penalty term on the norm of the constraints, and a step size on the update of the LMs. Therefore, it can affect both the optimality and the satisfaction of constraints. Fig.~\ref{fig:residuals_rho} shows the residual values as a function of a constant $\rho$, without using the adaptive $\rho$ update of \eqref{eq:adaptive_rho}, after 80 iterations. We notice that as $\rho$ increases, the value of the primal residual decreases because there is a higher penalty for the violation of the DSO constraints, whereas the value of the dual residual increases as the algorithm takes larger steps, which affects the optimal condition. A starting value of around 1 is an appropriate choice, as both residuals are below $10^{-4}$. The value is then adapted according to \eqref{eq:adaptive_rho} with values of $\tau^{incr},\tau^{decr}$ close to 1 if the two residuals diverge. 

\begin{figure}[!t]
    \centering
    \includegraphics[width=\linewidth]{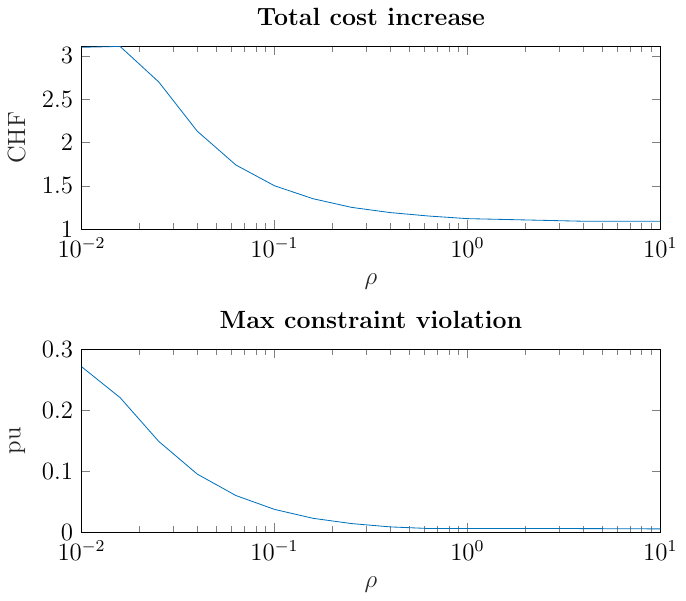}
    \caption{Effect of $\rho$ as a penalty term on the optimal solution obtained after 80 iterations of ADMM.}
    \label{fig:fees_rho}
\end{figure}

As seen in Eq.~\eqref{eq:prosumer_objective_rewriten}, the penalty term $\rho$ also influences the price signal advertised by the DSO. To quantify this effect, we first solve ADMM to convergence using the adaptive $\rho$ strategy of Section~\ref{sec:implementation}. Then, we minimize the price signal of Eq.~\eqref{eq:prosumer_objective_rewriten} for each prosumer using different values of $\rho$ and measure both the total increase in prosumer cost and the maximum constraint violation according to the prosumers' decisions. The results are shown in Fig.~\ref{fig:fees_rho}. We observe that both the total fees paid by the DSO and the constraint violation are minimal when $\rho$ exceeds approximately 1. Indeed $\rho$ does not affect the optimal value of the cost function because the term $||A_i \mathbf{x}_i-\mathbf{z}_i||^2_2$ in the Lagrangian (see Eq.~\eqref{eq:ci_definition}) is zero provided that the LMs $\mathbf{y}_i$ and the copied variables $\mathbf{z}_i$ have converged to their respective optimal values; it only influences how costs are split between the demand-dependent price and the constant fees (see Section~\ref{sec:admm_interpretation_new}). 
If $\rho \rightarrow 0$, the cost reduces to $(c+A_i^\top\mathbf{y}^k_i)^\top\mathbf{x}_i-(\mathbf{y}^k_i)^\top\mathbf{z}^k_i$, which no longer enforces the constraint \eqref{eq:prosumer_z_constraints}. In this case, if $\mathbf{y}_i$ has not fully converged, the prosumer’s decision may violate grid constraints or deviate from optimality. We thus conclude that $\rho$ does not affect the optimal solution, provided it is sufficiently large. However, the adaptive $\rho$ strategy might need many iterations to converge to an appropriate value of $\rho$, depending on the starting value $\rho^0$ and the values of $\tau^{incr}$ and $\tau^{decr}$. Therefore, a preliminary analysis with different values of $\rho$ is required to decide what constitutes a sufficiently large starting value $\rho^0$ before running ADMM with the adaptive strategy to fine-tune $\rho$.

\subsection{Experimental Results} \label{sec:experiments_results}

We test the proposed day-ahead strategy and real-time control using the experimental setup described at the beginning of this section to verify whether the performance with real hardware matches expectations.
The experiment spans the time frame from 10 AM to 7 PM, during which violations of the grid constraints are most likely. As the real-time control requires short-term forecasts of the PV generation (see Section~\ref{sec:rt_control}), a persistent predictor is used because, for short-time intervals, it is known to provide the best estimates \cite{scolari2016irradiance}. Each time the decisions of the prosumers are modified according to the receding horizon problem (see Section~\ref{sec:mpc}), the first 10 minutes of the day-ahead forecasts of the loads and the solar irradiance in the optimization problem are also modified using a persistent predictor.

\begin{figure}[t!]
  \centering
  \subfloat[]{%
        \includegraphics[width=1\linewidth]{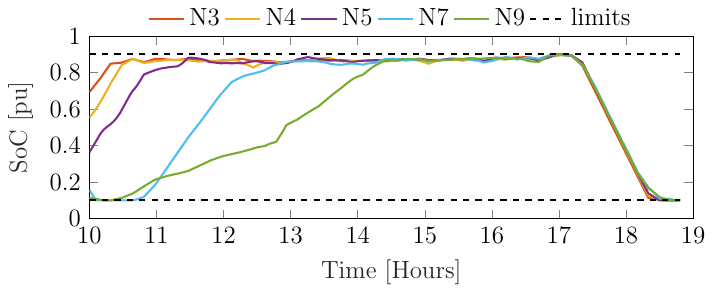}}
  \\
  \subfloat[]{\label{fig:experiments_pv}
        \includegraphics[width=1\linewidth]{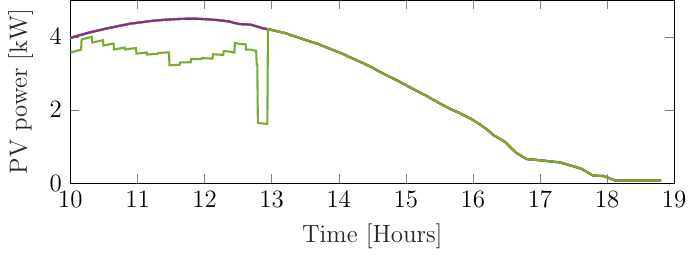}}
  \\
  \subfloat[]{%
        \includegraphics[width=1\linewidth]{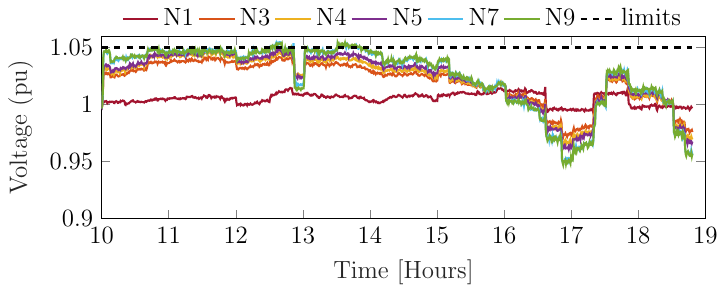}}
  \caption{Experimental results with emulated prosumers, equipped with simulated BESS and PV. (a) SoC of the BESSs, (b) PV production, and (c) measured nodal voltage magnitude, all of which fall within limits.}
  \label{fig:experiments_results}
\end{figure}

The experimental results are shown in Fig.~\ref{fig:experiments_results}. Data points refer to average power values within a 30-second time interval. As expected, the BESSs start charging in the middle of the day when the price is low and sell all their energy in the evening when the prosumer price is high. The nodal voltages are consistently within the voltage limits, verifying the MPC's ability to regulate the voltages despite the grid's uncertainties and the fact that the real-time control is unaware of the grid's topology. Moreover, we observe that the PV of the prosumer at N9 had to be curtailed, as expected in the day-ahead scheduling of Fig.~\ref{fig:results_coord} to respect the voltage constraint.

However, there are some discrepancies in the PV curtailment and the nodal voltages from the ones expected day-ahead. First of all, the slack voltage is not constant, as it is dictated by the upper grid. This causes some minor violations of the voltage upper bound at around 12:30 PM, which are corrected by the intra-day MPC in the following 10-minute cycle. In fact, it appears that the control decision for the cycle preceding 1 PM is more conservative, as the voltages dropped more than necessary. This happens because the PV curtailment of node N9 is higher than expected day-ahead (compare Fig.~\ref{fig:experiments_pv} to Fig.~\ref{fig:coord_pv}). This result is due to ADMM not converging at an acceptable tolerance within the required time constraints, which results in higher costs for the system. In the next cycle, the coordination is rerun so the DSO can correct the decisions of the prosumers with new price signals that respect the nodal voltages as much as necessary. This means that in a real setting with limited time constraints and uncertainty, it is possible that either the grid constraints are violated or the price signals are suboptimal within two computation cycles of the intra-day control. The duration of the suboptimal operation can be reduced by running the intra-day control more often; however, this would pose tighter constraints for ADMM to converge. Therefore, there is a trade-off between the frequency with which the coordination is run and the time we allow for the coordination to finish.

Nevertheless, the experiment demonstrates that the proposed algorithm can satisfy the grid's constraints while addressing the objectives of the prosumers. In addition, it confirms that the proposed methodology can effectively utilize a dynamic price signal to steer the flexibility of prosumers in providing voltage support to the grid. By re-applying coordination between the DSO and prosumers in a receding horizon, voltage constraints are satisfied despite forecasting uncertainties, provided that we leave enough time for the coordination phase to converge.

\section{Conclusions}
\label{sec:conclusion}
We have developed an algorithm to calculate price signals that encourage flexible prosumers to adjust their demand to respect the operational constraints of a distribution grid. The algorithm can be adopted by a DSO to dynamically modify an existing tariff, such as a ToU tariff or a static retail price, in real-time and prevent grid congestion and voltage violations by shifting price-sensitive loads. The algorithm is based on an interpretation of ADMM, enabling the analytical computation of a dynamic price signal through an iterative process. This process determines a price signal that minimizes prosumers' electricity bills while ensuring compliance with grid constraints. By leveraging the distributed nature of the ADMM framework, which solves local problems rather than relying on a centralized formulation, the algorithm maintains a clear separation between the DSO and prosumers. This avoids the need to exchange sensitive information such as individual flexibility levels or internal grid data.

Experiments are carried out in an LV distribution grid with nine nodes and five flexible prosumers, each equipped with a behind-the-meter BESS and curtailable PV generation.
The experiments demonstrate that the proposed algorithm effectively steers flexible prosumers in a practical setting, while a series of simulations illustrates convergence performance.
Future work will focus on incorporating uncertainty into the price signals, particularly in modeling generalized flexible resources, exploring effective strategies to address it, and addressing the issue of fairness that arises from advertising different price signals to the prosumers.

\bibliographystyle{IEEEtran}
\bibliography{references}

\end{document}